\documentclass[epj,twocolumn]{webofc}
\usepackage[varg]{txfonts}   
\usepackage{graphicx}

\wocname{epj}
\woctitle{Seismology of the Sun and the Distant Stars 2016}
\begin{document}
\title{Stellar magnetic activity and exoplanets}
%

\author{\firstname{A.~A.} \lastname{Vidotto}\inst{1}\fnsep\thanks{\email{Aline.Vidotto@tcd.ie}} 
}

\institute{School of Physics, Trinity College Dublin, the University of Dublin, Dublin-2, Ireland
          }

\abstract{%
It has been proposed that magnetic activity could be enhanced due to interactions between close-in massive planets and their host stars. In this article, I present a brief overview of the connection between stellar magnetic activity and exoplanets. 
Stellar activity can be probed in chromospheric lines, coronal emission, surface spot coverage, etc. Since these are manifestations of stellar magnetism, these measurements are often used as proxies for the magnetic field of stars. Here, instead of focusing on the magnetic proxies, I overview some recent results of magnetic field measurements using spectropolarimetric observations. Firstly, I discuss the general trends found between large-scale magnetism, stellar rotation, and coronal emission and show that magnetism seems to be correlated to the internal structure of the star. Secondly, I overview some works that show evidence that exoplanets could (or not) act as to enhance the activity of their host stars. 
}
\maketitle
%
\section{Introduction}\label{sec.intro}
Stellar activity can manifest itself in the form of, e.g.,  surface spot coverage, emissions from the chromosphere, transition region and corona \citep[e.g.,][]{1972ApJ...171..565S,1984ApJ...279..763N}. These manifestations are all indicators of  magnetic activity and are, therefore, often used as proxies for stellar magnetism. As stellar magnetism is responsible for driving stellar winds, flares and coronal mass ejections, stellar magnetic fields are the main driver of space weather on (exo)planets. Therefore, understanding the host star magnetism is a key step towards understanding how stellar activity affects exoplanets.

The host star magnetism can have effects on planetary mass-loss \citep{2012EP&S...64..179L,2014MNRAS.444.3761O}, habitability \citep{2013A&A...557A..67V, 2016A&A...596A.111R}, and on the magnetospheres of exoplanets \citep{2014A&A...570A..99S}, to name a few. It has also been suggested that exoplanets, in particular those orbiting close to the star, can also affect stellar activity, through tidal interactions and/or magnetic reconnection events \citep{2000ApJ...533L.151C, 2005ApJ...622.1075S, 2009A&A...505..339L,2014A&A...565L...1P,2016A&A...593A.128P}.

In this article, I present a brief review of stellar surface magnetism (Section \ref{sec.mag}) and some proposed effects that exoplanets might have on the activity of their host stars (Section \ref{sec.pla}).

\section{Stellar surface  magnetism: brief overview of latest results}\label{sec.mag}
Magnetic fields are responsible for driving the winds of cool, main-sequence stars. These outflows carry away angular momentum from the star, slowing down its rotation. Since rotation is intimately related to the generation of stellar magnetic fields, with age, stars spin down and the magnetic fields they generate decrease in magnitude \citep{2014MNRAS.441.2361V}. 

One way to map surface magnetism is through the tomographic imaging technique Zeeman Doppler imaging (ZDI) \citep{1997A&A...326.1135D}. This technique has now successfully mapped the large-scale surface field of hundreds of cool, pre-main-sequence and main-sequence stars \citep[e.g.,][]{2009ARA&A..47..333D}. Due to the increasing sample of stars with mapped fields, several trends are emerging \citep{2008MNRAS.388...80P, 2009ARA&A..47..333D, 2010MNRAS.407.2269M, 2015MNRAS.453.4301S, 2014MNRAS.441.2361V, 2016MNRAS.461.1465H}. For example, it has been shown that the topology of surface magnetic fields depends on the internal structure, where fully convective stars present mainly poloidal fields \citep{2010MNRAS.407.2269M, 2015MNRAS.453.4301S} as compared to stars that have a radiative core. This change in topology might be caused by the development of the tachocline \citep{2010MNRAS.407.2269M, 2015MNRAS.453.4301S}.

\begin{figure*}[t]
\centering
\includegraphics[width=\hsize]{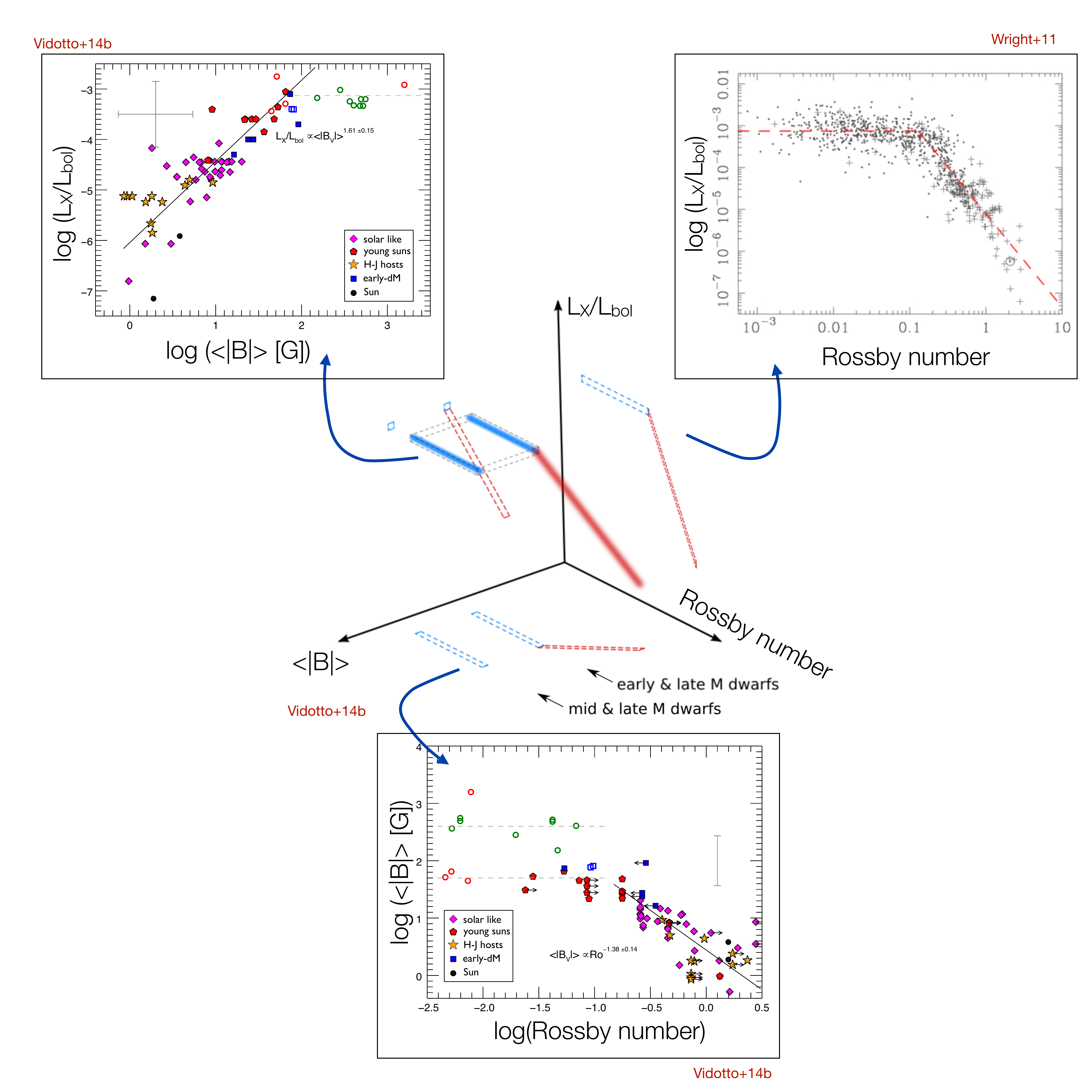}
\caption{Stellar activity is a complex function of many variables, such as age, mass, rotation, and magnetism. This figure illustrates how magnetic fields are related to X-ray emission (left-most panel) and to Rossby number (bottom panel). It also shows the more traditional activity-rotation relation between X-ray emission and Rossby number (right-most panel). Each of these panels are projections of a 3D distribution of $\{L_x/L_{\rm bol}, {\rm Ro}, {\langle |B| \rangle} \}$ (middle panel). This figure is based on the plots presented in \cite{2014MNRAS.441.2361V} and \cite{2011ApJ...743...48W}.}
\label{fig-1}       
\end{figure*}

Evolutionary speaking, magnetic fields of main-sequence stars are also observed to decrease with age $t$. \cite{2014MNRAS.441.2361V} showed that the average magnetic field intensity ${\langle |B| \rangle}$ is a function of $t^{-0.66\pm 0.05}$. This age decay has a power law consistent to the power-law of $-0.5$ between rotation rate $\Omega_\star$ and age \citep{1972ApJ...171..565S}: $\Omega_\star \propto t^{-1/2}$. This similarity could be an indication that the dynamo operating inside the star is of a linear type (i.e., the magnetic field is linearly proportional to stellar rotation rate), as already suggested by \cite{1972ApJ...171..565S}. Indeed, ZDI observations have revealed that, for stars with Rossby numbers Ro $\gtrsim 0.1$, the average large-scale magnetic field decays with Ro$^{-1.38\pm 0.14}$ \citep{2014MNRAS.441.2361V} (see bottom panel in Figure \ref{fig-1}). Since Rossby number is defined as Ro $=2\pi (\Omega_\star \tau_{conv})^{-1}$, where $\tau_{conv}$ is the convective turnover time, this relation indicates an approximately linear decay of magnetism with rotation rate.\footnote{A decay is also observed between magnetism and Rossby number for active evolved (single) G -- K giants, albeit with a different slope: \cite{2015A&A...574A..90A} showed that the line-of-sight magnetic field strength $B_l$ decays with Ro$^{-0.68}$. A recent study proposes that planet engulfment could be enhancing the magnetic field of giant stars \citep{2016A&A...593L..15P}. I will return to this point in section \ref{sec.pla}.}

Stars with Ro $\lesssim 0.1$ are {\it saturated}, i.e., their activity does not vary with Rossby number. This saturation has been identified, for example, in the relations between $L_x/L_{\rm bol}$ and Ro \citep{2003A&A...397..147P,2011MNRAS.411.2099J, 2011ApJ...743...48W}, where $L_x$ is the X-ray luminosity and $L_{\rm bol}$ the stellar bolometric luminosity  (see right-most panel in Figure \ref{fig-1}). When investigating the behaviour of stellar magnetism for a sample of saturated early- and mid-M dwarf stars, \citep{2008MNRAS.390..545D,2008MNRAS.390..567M} noticed that these stars saturated at different magnetic field strengths, whereby early-M dwarfs presented a lower saturation threshold than mid-M dwarfs  (see dashed lines in the bottom panel in Figure \ref{fig-1}).  These authors suggested that this could be caused by different dynamo efficiencies at producing small- and large-scale magnetic fields. 

Since rotation is linked to magnetic field strength and X-ray emission is linked to rotation, then the immediate question we asked ourselves is how magnetic field strength can be related to X-ray emission. The left-most panel of Figure \ref{fig-1} shows precisely this investigation, where we note that X-ray emission increases with average magnetic field strength ${\langle |B| \rangle} ^{1.61\pm 0.15}$. These three diagrams, illustrated in Figure \ref{fig-1}, are actually projections of a 3D distribution of $\{L_x/L_{\rm bol}, {\rm Ro}, {\langle |B| \rangle} \}$ (sketch shown in the middle panel). In fact, stellar activity is a function of many other variables, such as age and stellar mass, with the sketch presented in Figure \ref{fig-1} representing a cut of a multi-dimensional function.

\begin{figure*}[t]
\centering
\includegraphics[width=0.49\hsize]{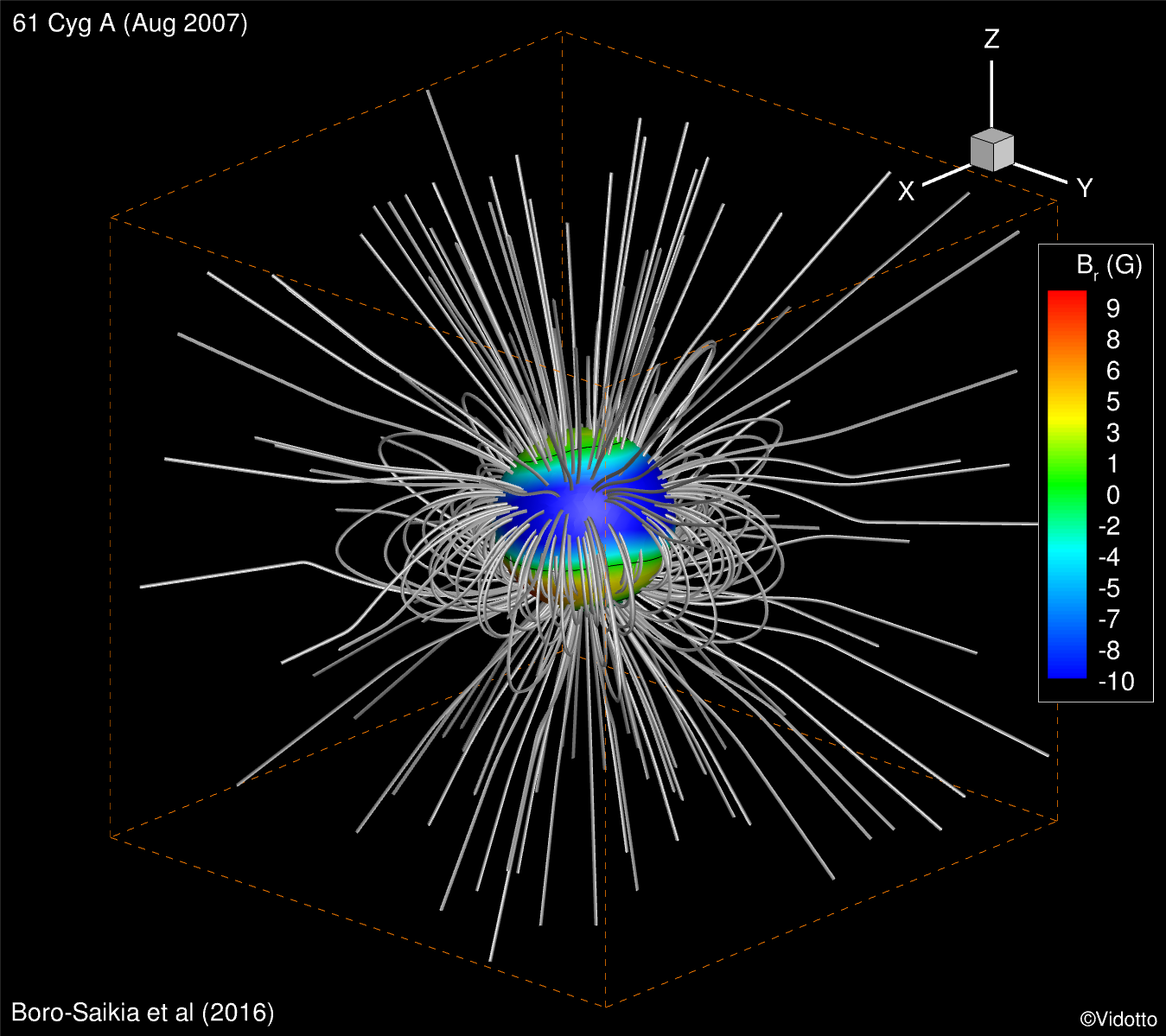}
\includegraphics[width=0.49\hsize]{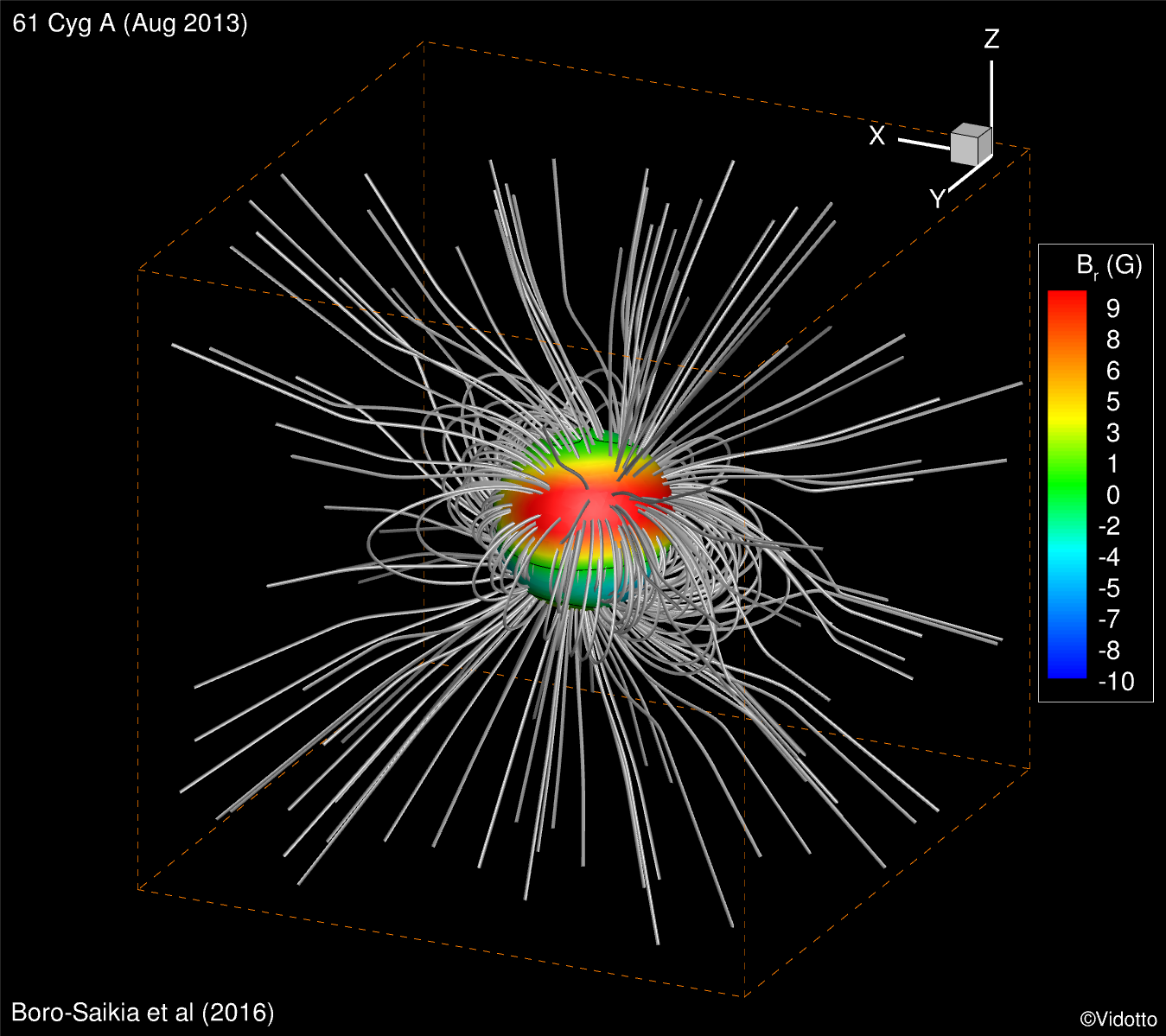}
\caption{The star 61 Cyg A shows polarity reversals on the large-scale field with a period of a full magnetic cycle (i.e., two reversals) of around 7 years. The reversals are in phase with the chromospheric and coronal cycle. Figure based on the ZDI maps of  \cite{2016A&A...594A..29B} for Aug 2007 (left) and Aug 2013 (right). The magnetic field lines are extrapolated using a potential field source surface method, which takes into account the radial component of the observed field from \cite{2016A&A...594A..29B}.}
\label{fig-2}       
\end{figure*}

A common characteristic of the plots regarding stellar activity are their relatively large spread. Part of this spread is due to intrinsic variability of stellar activity and not only due to observational uncertainties. The Sun, for instance, varies in magnetic field intensity and X-ray emission along its cycle  (see left-most and bottom panels of Figure \ref{fig-1}, showing two points for the Sun, one during activity minimum and the other during maximum of activity). Other stars present, like our Sun, variations on their surface magnetic fields, some of which are associated to well-defined magnetic cycles \citep{2008MNRAS.385.1179D,2009MNRAS.398.1383F,2016A&A...594A..29B,2016MNRAS.459.4325M}. This medium term variability (on the order of years) acts as to increase the spread in rotation-magnetic-activity relations. Figure \ref{fig-2} illustrates the case of 61 Cyg A \citep{2016A&A...594A..29B}. This star shows polarity reversals on the large-scale field with a period of a full magnetic cycle (i.e., two reversals) of around 7 years. The reversals are in phase with the chromospheric and coronal cycle, similar to what is observed for the Sun.

\section{Can planets influence the activity of their host stars? }\label{sec.pla}
That stellar activity could affect exoplanets is not hard to imagine. After all, the Earth and the planets in the solar system regularly experience the effects of solar energetic particles interacting with their magnetospheres/atmospheres. Several works have studied the effects of the stars and their winds on their surrounding exoplanets \citep[e.g.][just to name a few]{2004ApJ...602L..53I,2007AsBio...7..167K,2008A&A...490..843J,2014ApJ...795...86S,2014A&A...570A..99S,2015A&A...577L...3T,2015MNRAS.449.4117V,2015ApJ...799L..15K,2016MNRAS.459.1907N}. But could the opposite also occur? Can planets influence the activity of their host stars?

Motivated by the increase in chromospheric and coronal activity observed in interacting binary stars, \cite{2000ApJ...533L.151C} proposed that close-in giant planets (or brown dwarfs) could lead to enhancement of dynamo activity. These planet-star interactions were proposed to be of two types: magnetospheric or tidal interactions. In the first case, reconnection events between the magnetic field lines of an exoplanet and of the host star would accelerate particles that would travel along stellar magnetic field lines, interacting with the outer layers of the stellar atmosphere. This interaction would then generate enhanced heat near the magnetic field footpoints. This anomalous activity would be modulated by the orbital period of the planet, as opposed to the `normal' stellar activity, which is modulated by the stellar rotation period. In the second case, tides raised on the star by the revolving planet would cause repeated expansion and contraction of  tidal bulges, generating waves that can dissipate and deposit energy in the upper atmosphere of the star (i.e., transition region, chromosphere). Like in the first scenario, this anomalous activity increase would also be cyclic, but the modulation would take place at half orbital period. 

 Searches for anomalous activity have since been on going. In a series of works, \cite{2003ApJ...597.1092S,2005ApJ...622.1075S} presented evidence of anomalous  activity on a few planet-hosting stars. These anomalies were modulated with the orbital period of their close-in giant planets, possibly indicating a signature of planet-star interaction. Repeated observations (at different epochs), however, failed to detect such modulations \citep{2008ApJ...676..628S}, leading the authors to suggest that activity enhancement is intermittent and could depend, among others, on the configuration of the star's magnetic field.  Searching for chromospheric/transition region signatures of planet-star interaction has developed into an active field of research, yielding positive (or tentatively positive) \citep{2012PASA...29..141G,2016ApJ...820...89F,2015ApJ...811L...2M} and negative \citep{2010MNRAS.406..409F,2012ApJ...754..137M,2013A&A...552A...7S,2011A&A...530A..73C,2016A&A...592A.143F} detections. 
 
 In a recent work, for instance, \cite{2015ApJ...811L...2M} conducted X-ray observations of HD 17156, the host of an eccentric planet, and reported enhanced X-ray stellar emission a few hours after the periastron passage. This emission, happening near the closest approach of the planet where planet-star interactions are expected to be enhanced, could have been mediated by magnetic reconnection or accretion of planetary material towards the star \cite[e.g.,][]{2015ApJ...805...52P}. Following the same idea, i.e., that eccentric planets can trigger stellar activity at orbital phases near the periastron passage, \cite{2016A&A...592A.143F} investigated if the activity level of HD 80606 could have been influenced by its eccentric planet. For that, they used optical spectroscopy and photometric monitoring of the star during the periastron passage and close to the apastron passage. They could not find any evidence that the stellar activity level was modified between these two observing epochs. An interpretation (the simplest one) is that this planet does not trigger activity enhancement on the star, or if does, this enhancement would be below the detection sensitivity. Another possibility is that the enhanced activity would appear with a certain phase lag outside the observing window.
 
If close-in exoplanets can indeed enhance stellar magnetic activity, it might also be possible to verify that in magnetic images of planet-hosting stars. Indeed, among the ZDI sample shown in Figure \ref{fig-1}, several of the stars are known to host close-in giant planets (i.e., hot-Jupiters). These stars are shown as yellow star symbols in Figure \ref{fig-1}. The close-in, massive planets are expected to cause the largest observational signature of its influence on the large-scale magnetic field of the host star. In terms of magnetic characteristics (topology and intensity), however, there has not been any measurable difference between the hot-Jupiter hosts and the remaining stars of the sample \citep{2013MNRAS.435.1451F, 2014MNRAS.441.2361V, matthew}. In other words, the yellow star symbols shown in Figure \ref{fig-1} do not seem to occupy a particular position in these diagrams with respect to other symbols in these plots (i.e., the stars not known to host hot Jupiters).

Another possibility is that the interaction between a close-in planet and the host star might be able to affect stellar {\it cycles}. The famous planet-host star $\tau$ Boo, has been monitored with ZDI for about a decade \cite{2007MNRAS.374L..42C,2008MNRAS.385.1179D,2009MNRAS.398.1383F,2013MNRAS.435.1451F,2016MNRAS.459.4325M}. This star shows flips in magnetic field polarity every year, making it the star with  the shortest full (2 years) magnetic cycle mapped with ZDI. Since $\tau$ Boo is a F7V star, its convective envelope is quite thin and it is estimated to have a mass similar to that of the hot Jupiter. The idea is that, through tidal interactions, the planet and the stellar convective envelope would have been synchronised, but the interior radiative core of the star would still be rotating at a different rate. The shear created by different rotation rates between the radiative interior and the convective envelope could be the cause of the short magnetic cycle. According to this interpretation, the planet would have indirectly contributed to altering the generation of magnetic fields in the star. An alternative interpretation involves the high level of surface differential rotation derived in ZDI studies \cite{2009MNRAS.398.1383F}.

In a series of works, \cite{2016A&A...591A..45P,2016A&A...593A.128P,2016A&A...593L..15P} proposed that planets (initially at large separations) could be engulfed by stars, when their hosts evolve off of the main sequence. With the expansion of the stars at the red giant branch, tidal interactions between a Jupiter-like planet and the star trigger planetary migration, leading the planet to  eventually be engulfed by the star. During the migration process, orbital angular momentum are transferred to stellar spin. Therefore, these red giant branch stars would present higher rotation rates, which could not be explained by any reasonable model for single star evolution \citep{2016A&A...591A..45P,2016A&A...593A.128P}. From the observational sample of \cite{2012ApJ...757..109C}, \cite{2016A&A...591A..45P,2016A&A...593A.128P,2016A&A...593L..15P} identified  some fast rotating red giants that are candidates of having engulfed their planets in a recent past. Since rotation and magnetic fields are closely related \citep{2015A&A...574A..90A}, \cite{2016A&A...593L..15P} then further proposed, that these fast rotating red giants would also show a higher magnetic field. This idea can, of course, be directly tested with spectropolarimetric observations. 

\section{Summary and Conclusions}
This article was based on the talk I presented at the conference ``Seismology of the Sun and the Distant Stars 2016'' on  the topic of ``stellar-planet activity''. First, I would like to remind the reader that the material I presented during my talk, and that was described in this article, is far from a comprehensive overview of the theme. I hope I could provide some starting references for an interested reader to proceed on their own. 

In particular, I had the (hard) task to discuss synergies between ``stellar-planet activity'' and asteroseismology. The first immediate synergy one can identify is on the relation between the internal structure of  stars (probed with asteroseismology) and the global properties of stellar magnetic fields (probed in ZDI studies, Section \ref{sec.mag}). In the second part of my talk, I focused on how exoplanets might or not enhance stellar magnetic activity. I showed that, although there is some evidence that planets can trigger magnetic activity on their host stars, this evidence seems to be intermittent, making it more difficult to precise the nature of the planet-star interaction. In fact, there is an on-going debate in the community regarding the detectability of  the signature of planet-star interaction. Would there be a way that asteroseismology could contribute to this debate?

\section*{Acknowledgements}
I would like to warmly thank the organisers of  ``Seismology of the Sun and the Distant Stars 2016'' for the financial support offered for me to attend this conference. Obrigada!


\def\aj{{AJ}}                   
\def\araa{{ARA\&A}}             
\def\apj{{ApJ}}                 
\def\apjl{{ApJ~Letters}}                
\def\apjs{{ApJS}}               
\def\apss{{Ap\&SS}}             
\def\aap{{A\&A}}                
\def\aapr{{A\&A~Rev.}}          
\def\aaps{{A\&AS}}              
\def\mnras{{MNRAS}}             
\def\mnrasl{{MNRAS~Letters}}             
\def\pasp{{PASP}}               
\def\solphys{{Sol.~Phys.}}      
\def\sovast{{Soviet~Ast.}}      
\def\ssr{{Space~Sci.~Rev.}}     
\def\nat{{Nature}}              
\def\iaucirc{{IAU~Circ.}}       
\def\planss{{Planetary Space Science}}
\def\jgr{{JGR}}   
\def\grl{{Geophysical Research Letters}}
\def\icarus{{Icarus}}
\def\jcp{{J.~of Comp.~Physics}}
\def\pasa{{Publications of the Astronomical Society of Australia}}
\let\astap=\aap
\let\apjlett=\apjl
\let\apjsupp=\apjs
\let\applopt=\ao


\begin{thebibliography}{100}

\bibitem{1972ApJ...171..565S}
A.~{Skumanich}, \apj \textbf{171}, 565 (1972)

\bibitem{1984ApJ...279..763N}
R.W. {Noyes}, L.W. {Hartmann}, S.L. {Baliunas}, D.K. {Duncan}, A.H. {Vaughan},
  \apj , \textbf{279}, 763 (1984)

\bibitem{2012EP&S...64..179L}
H.~{Lammer}, M.~{Guedel}, Y.~{Kulikov}, I.~{Ribas}, T.V. {Zaqarashvili}, M.L.
  {Khodachenko}, K.G. {Kislyakova}, H.~{Gr{\"o}ller}, P.~{Odert},
  M.~{Leitzinger} et~al., Earth, Planets, and Space,  \textbf{64}, 179 (2012)

\bibitem{2014MNRAS.444.3761O}
J.E. {Owen}, F.C. {Adams}, \mnras , \textbf{444}, 3761 (2014)

\bibitem{2013A&A...557A..67V}
A.A. {Vidotto}, M.~{Jardine}, J.~{Morin}, J.F. {Donati}, P.~{Lang}, A.J.B.
  {Russell}, \aap , \textbf{557}, A67 (2013)

\bibitem{2016A&A...596A.111R}
I.~{Ribas}, E.~{Bolmont}, F.~{Selsis}, A.~{Reiners}, J.~{Leconte}, S.N.
  {Raymond}, S.G. {Engle}, E.F. {Guinan}, J.~{Morin}, M.~{Turbet} et~al., \aap ,
  \textbf{596}, A111 (2016)

\bibitem{2014A&A...570A..99S}
V.~{See}, M.~{Jardine}, A.A. {Vidotto}, P.~{Petit}, S.C. {Marsden}, S.V.
  {Jeffers}, J.D. {do Nascimento}, \aap , \textbf{570}, A99 (2014)

\bibitem{2000ApJ...533L.151C}
M.~{Cuntz}, S.H. {Saar}, Z.E. {Musielak}, \apjl , \textbf{533}, L151 (2000)

\bibitem{2005ApJ...622.1075S}
E.~{Shkolnik}, G.A.H. {Walker}, D.A. {Bohlender}, P.G. {Gu}, M.~{K{\"u}rster},
  \apj , \textbf{622}, 1075 (2005)

\bibitem{2009A&A...505..339L}
A.F. {Lanza}, \aap , \textbf{505}, 339 (2009)

\bibitem{2014A&A...565L...1P}
K.~{Poppenhaeger}, S.J. {Wolk}, \aap , \textbf{565}, L1 (2014)

\bibitem{2016A&A...593A.128P}
G.~{Privitera}, G.~{Meynet}, P.~{Eggenberger}, A.A. {Vidotto}, E.~{Villaver},
  M.~{Bianda}, \aap , \textbf{593}, A128 (2016)

\bibitem{2014MNRAS.441.2361V}
A.A. {Vidotto}, S.G. {Gregory}, M.~{Jardine}, J.F. {Donati}, P.~{Petit},
  J.~{Morin}, C.P. {Folsom}, J.~{Bouvier}, A.C. {Cameron}, G.~{Hussain} et~al.,
  \mnras , \textbf{441}, 2361 (2014)

\bibitem{1997A&A...326.1135D}
J.F. {Donati}, S.F. {Brown}, \aap , \textbf{326}, 1135 (1997)

\bibitem{2009ARA&A..47..333D}
J.~{Donati}, J.D. {Landstreet}, \araa , \textbf{47}, 333 (2009)

\bibitem{2008MNRAS.388...80P}
P.~{Petit}, B.~{Dintrans}, S.K. {Solanki}, J.F. {Donati}, M.~{Auri{\`e}re},
  F.~{Ligni{\`e}res}, J.~{Morin}, F.~{Paletou}, J.~{Ramirez Velez}, C.~{Catala}
  et~al., \mnras , \textbf{388}, 80 (2008)

\bibitem{2010MNRAS.407.2269M}
J.~{Morin}, J.~{Donati}, P.~{Petit}, X.~{Delfosse}, T.~{Forveille}, M.M.
  {Jardine}, \mnras , \textbf{407}, 2269 (2010)

\bibitem{2015MNRAS.453.4301S}
V.~{See}, M.~{Jardine}, A.A. {Vidotto}, J.F. {Donati}, C.P. {Folsom}, S.~{Boro
  Saikia}, J.~{Bouvier}, R.~{Fares}, S.G. {Gregory}, G.~{Hussain} et~al.,
  \mnras , \textbf{453}, 4301 (2015)

\bibitem{2016MNRAS.461.1465H}
{\'E}.M. {H{\'e}brard}, J.F. {Donati}, X.~{Delfosse}, J.~{Morin}, C.~{Moutou},
  I.~{Boisse}, \mnras , \textbf{461}, 1465 (2016)

\bibitem{2015A&A...574A..90A}
M.~{Auri{\`e}re}, R.~{Konstantinova-Antova}, C.~{Charbonnel}, G.A. {Wade},
  S.~{Tsvetkova}, P.~{Petit}, B.~{Dintrans}, N.A. {Drake}, T.~{Decressin},
  N.~{Lagarde} et~al., \aap , \textbf{574}, A90 (2015)

\bibitem{2016A&A...593L..15P}
G.~{Privitera}, G.~{Meynet}, P.~{Eggenberger}, C.~{Georgy}, S.~{Ekstr{\"o}m},
  A.A. {Vidotto}, M.~{Bianda}, E.~{Villaver}, A.~{ud-Doula}, \aap , \textbf{593},
  L15 (2016)

\bibitem{2011ApJ...743...48W}
N.J. {Wright}, J.J. {Drake}, E.E. {Mamajek}, G.W. {Henry}, \apj , \textbf{743},
  48 (2011)

\bibitem{2003A&A...397..147P}
N.~{Pizzolato}, A.~{Maggio}, G.~{Micela}, S.~{Sciortino}, P.~{Ventura}, \aap ,
  \textbf{397}, 147 (2003)

\bibitem{2011MNRAS.411.2099J}
R.D. {Jeffries}, R.J. {Jackson}, K.R. {Briggs}, P.A. {Evans}, J.P. {Pye},
  \mnras , \textbf{411}, 2099 (2011)

\bibitem{2008MNRAS.390..545D}
J.~{Donati}, J.~{Morin}, P.~{Petit}, X.~{Delfosse}, T.~{Forveille},
  M.~{Auri{\`e}re}, R.~{Cabanac}, B.~{Dintrans}, R.~{Fares}, T.~{Gastine}
  et~al., \mnras , \textbf{390}, 545 (2008)

\bibitem{2008MNRAS.390..567M}
J.~{Morin}, J.~{Donati}, P.~{Petit}, X.~{Delfosse}, T.~{Forveille},
  L.~{Albert}, M.~{Auri{\`e}re}, R.~{Cabanac}, B.~{Dintrans}, R.~{Fares}
  et~al., \mnras , \textbf{390}, 567 (2008)

\bibitem{2008MNRAS.385.1179D}
J.F. {Donati}, C.~{Moutou}, R.~{Far{\`e}s}, D.~{Bohlender}, C.~{Catala},
  M.~{Deleuil}, E.~{Shkolnik}, A.~{Collier Cameron}, M.M. {Jardine}, G.A.H.
  {Walker}, \mnras , \textbf{385}, 1179 (2008)

\bibitem{2009MNRAS.398.1383F}
R.~{Fares}, J.~{Donati}, C.~{Moutou}, D.~{Bohlender}, C.~{Catala},
  M.~{Deleuil}, E.~{Shkolnik}, A.C. {Cameron}, M.M. {Jardine}, G.A.H. {Walker},
  \mnras , \textbf{398}, 1383 (2009)

\bibitem{2016A&A...594A..29B}
S.~{Boro Saikia}, S.V. {Jeffers}, J.~{Morin}, P.~{Petit}, C.P. {Folsom}, S.C.
  {Marsden}, J.F. {Donati}, R.~{Cameron}, J.C. {Hall}, V.~{Perdelwitz} et~al.,
  \aap , \textbf{594}, A29 (2016)

\bibitem{2016MNRAS.459.4325M}
M.W. {Mengel}, R.~{Fares}, S.C. {Marsden}, B.D. {Carter}, S.V. {Jeffers},
  P.~{Petit}, J.F. {Donati}, C.P. {Folsom}, {BCool Collaboration}, \mnras ,
  \textbf{459}, 4325 (2016)

\bibitem{2004ApJ...602L..53I}
W.H. {Ip}, A.~{Kopp}, J.H. {Hu}, \apjl , \textbf{602}, L53 (2004)

\bibitem{2007AsBio...7..167K}
M.L. {Khodachenko}, I.~{Ribas}, H.~{Lammer}, J.~{Grie{\ss}meier}, M.~{Leitner},
  F.~{Selsis}, C.~{Eiroa}, A.~{Hanslmeier}, H.K. {Biernat}, C.J. {Farrugia}
  et~al., Astrobiology , \textbf{7}, 167 (2007)

\bibitem{2008A&A...490..843J}
M.~{Jardine}, A.C. {Cameron}, \aap , \textbf{490}, 843 (2008)

\bibitem{2014ApJ...795...86S}
A.~{Strugarek}, A.S. {Brun}, S.P. {Matt}, V.~{R{\'e}ville}, \apj , \textbf{795},
  86 (2014)

\bibitem{2015A&A...577L...3T}
L.~{Tu}, C.P. {Johnstone}, M.~{Guedel}, H.~{Lammer}, \aap , \textbf{577}, L3
  (2015)

\bibitem{2015MNRAS.449.4117V}
A.A. {Vidotto}, R.~{Fares}, M.~{Jardine}, C.~{Moutou}, J.F. {Donati}, \mnras ,
  \textbf{449}, 4117 (2015)

\bibitem{2015ApJ...799L..15K}
K.G. {Kislyakova}, L.~{Fossati}, C.P. {Johnstone}, M.~{Holmstr{\"o}m}, V.V.
  {Zaitsev}, H.~{Lammer}, \apjl  , \textbf{799}, L15 (2015)

\bibitem{2016MNRAS.459.1907N}
B.A. {Nicholson}, A.A. {Vidotto}, M.~{Mengel}, L.~{Brookshaw}, B.~{Carter},
  P.~{Petit}, S.C. {Marsden}, S.V. {Jeffers}, R.~{Fares}, {the BCool
  Collaboration}, \mnras , \textbf{459}, 1907 (2016)

\bibitem{2003ApJ...597.1092S}
E.~{Shkolnik}, G.A.H. {Walker}, D.A. {Bohlender}, \apj , \textbf{597}, 1092
  (2003)

\bibitem{2008ApJ...676..628S}
E.~{Shkolnik}, D.A. {Bohlender}, G.A.H. {Walker}, A.~{Collier Cameron}, \apj ,
  \textbf{676}, 628 (2008)

\bibitem{2012PASA...29..141G}
L.~{Gurdemir}, S.~{Redfield}, M.~{Cuntz}, \pasa , \textbf{29}, 141 (2012)

\bibitem{2016ApJ...820...89F}
K.~{France}, R.O. {Parke Loyd}, A.~{Youngblood}, A.~{Brown}, P.C. {Schneider},
  S.L. {Hawley}, C.S. {Froning}, J.L. {Linsky}, A.~{Roberge}, A.P. {Buccino}
  et~al., \apj , \textbf{820}, 89 (2016)

\bibitem{2015ApJ...811L...2M}
A.~{Maggio}, I.~{Pillitteri}, G.~{Scandariato}, A.F. {Lanza}, S.~{Sciortino},
  F.~{Borsa}, A.S. {Bonomo}, R.~{Claudi}, E.~{Covino}, S.~{Desidera} et~al.,
  \apjl , \textbf{811}, L2 (2015)

\bibitem{2010MNRAS.406..409F}
R.~{Fares}, J.~{Donati}, C.~{Moutou}, M.M. {Jardine}, J.~{Grie{\ss}meier},
  P.~{Zarka}, E.L. {Shkolnik}, D.~{Bohlender}, C.~{Catala}, A.C. {Cameron},
  \mnras , \textbf{406}, 409 (2010)

\bibitem{2012ApJ...754..137M}
B.P. {Miller}, E.~{Gallo}, J.T. {Wright}, A.K. {Dupree}, \apj , \textbf{754}, 137
  (2012)

\bibitem{2013A&A...552A...7S}
G.~{Scandariato}, A.~{Maggio}, A.F. {Lanza}, I.~{Pagano}, R.~{Fares}, E.L.
  {Shkolnik}, D.~{Bohlender}, A.C. {Cameron}, S.~{Dieters}, J.F. {Donati}
  et~al., \aap , \textbf{552}, A7 (2013)

\bibitem{2011A&A...530A..73C}
B.L. {Canto Martins}, M.L. {Das Chagas}, S.~{Alves}, I.C. {Le{\~a}o}, L.P. {de
  Souza Neto}, J.R. {de Medeiros}, \aap , \textbf{530}, A73 (2011)

\bibitem{2016A&A...592A.143F}
P.~{Figueira}, A.~{Santerne}, A.~{Su{\'a}rez Mascare{\~n}o}, J.~{Gomes da
  Silva}, L.~{Abe}, V.Z. {Adibekyan}, P.~{Bendjoya}, A.C.M. {Correia},
  E.~{Delgado-Mena}, J.P. {Faria} et~al., \aap , \textbf{592}, A143 (2016)

\bibitem{2015ApJ...805...52P}
I.~{Pillitteri}, A.~{Maggio}, G.~{Micela}, S.~{Sciortino}, S.J. {Wolk},
  T.~{Matsakos}, \apj , \textbf{805}, 52 (2015)

\bibitem{2013MNRAS.435.1451F}
R.~{Fares}, C.~{Moutou}, J.F. {Donati}, C.~{Catala}, E.L. {Shkolnik}, M.M.
  {Jardine}, A.C. {Cameron}, M.~{Deleuil}, \mnras , \textbf{435}, 1451 (2013)

\bibitem{matthew}
M.W. {Mengel}, S.C. {Marsden}, B.D. {Carter}, J.~{Horner}, R.~{King},
  R.~{Fares}, S.V. {Jeffers}, P.~{Petit}, A.A. {Vidotto}, {BCool
  Collaboration}, \mnras , \textbf{465}, 2734, (2017)

\bibitem{2007MNRAS.374L..42C}
C.~{Catala}, J.F. {Donati}, E.~{Shkolnik}, D.~{Bohlender}, E.~{Alecian},
  \mnrasl , \textbf{374}, L42 (2007)

\bibitem{2016A&A...591A..45P}
G.~{Privitera}, G.~{Meynet}, P.~{Eggenberger}, A.A. {Vidotto}, E.~{Villaver},
  M.~{Bianda}, \aap , \textbf{591}, A45 (2016)

\bibitem{2012ApJ...757..109C}
J.K. {Carlberg}, K.~{Cunha}, V.V. {Smith}, S.R. {Majewski}, \apj , \textbf{757},
  109 (2012)

\end{thebibliography}
\end{document}